\documentclass{article}
\pdfoutput=1
\usepackage{pst-pdf}
\usepackage{amsmath,amsfonts}
\usepackage{subfigure}
\graphicspath{{./}{figs/}}
\usepackage{graphicx}
\usepackage{psfrag}
\usepackage{dcolumn}
\usepackage{rotating}
\usepackage{xspace}
\usepackage{url}
\usepackage{upgreek}
\usepackage{multicol}
\usepackage{pict2e,slashbox,multirow}
\usepackage{verbatim}
\usepackage[hmargin=1.0in]{geometry}
\usepackage{yhmath}

\begin{document}
\title{Shock Dynamics in Layered Periodic Media}

\author{%
  David I. Ketcheson\thanks{Mathematical and Computer Sciences Division, King Abdullah University of Science and
    Technology, 4700 KAUST, Thuwal 23955, Saudi Arabia.
   (\protect\url{david.ketcheson@kaust.edu.sa})} \and
    Randall J. LeVeque\thanks{Department of Applied Mathematics, University of Washington, Box 352420, Seattle, WA 98195-2420.
   (\mbox{rjl@uw.edu}) }}

\maketitle

\bibliographystyle{siam}

\begin{abstract}
Solutions of constant-coefficient nonlinear hyperbolic PDEs generically 
develop shocks, even if the initial data is smooth.  Solutions of hyperbolic
PDEs with variable coefficients can behave very differently.
We investigate formation and stability of shock waves in a one-dimensional 
periodic layered medium by computational study of time-reversibility and entropy 
evolution.
We find that periodic layered media tend to inhibit shock formation.
For small initial conditions and large impedance variation, 
no shock formation is detected even
after times much greater than the time of shock formation in a homogeneous
medium.  Furthermore, weak shocks are observed to be dynamically unstable in
the sense that they do not lead to significant long-term entropy decay.
We propose a characteristic condition for admissibility of shocks in heterogeneous
media that generalizes the classical Lax entropy condition
and accurately predicts the formation or absence of shocks in these media.
\end{abstract}

\pagestyle{myheadings}
\thispagestyle{plain}
\markboth{D. I. Ketcheson and R. J. LeVeque}{Shock dynamics in periodic media}



\newcommand{\qh}{\hat{q}}
\newcommand{\be}{\begin{equation}}
\newcommand{\ee}{\end{equation}}
\newcommand{\bq}{\mathbf{q}}
\newcommand{\bx}{\mathbf{x}}
\newcommand{\br}{\mathbf{r}}
\newcommand{\imh}{{i-\frac{1}{2}}}
\newcommand{\iph}{{i+\frac{1}{2}}}
\newcommand{\ipmh}{{i \pm \frac{1}{2}}}
\newcommand{\jph}{{j+\frac{1}{2}}}
\newcommand{\Aop}{{\cal A}}
\newcommand{\Bop}{{\cal B}}
\newcommand{\Wop}{{\cal W}}
\newcommand{\Oop}{{\cal O}}
\newcommand{\DQ}{\Delta Q}
\newcommand{\Dq}{\Delta q}
\newcommand{\Dx}{\Delta x}
\newcommand{\Dy}{\Delta y}
\newcommand{\Du}{\Delta u}
\newcommand{\bu}{\mathbf{u}}
\newcommand{\bv}{\mathbf{v}}
\newcommand{\bw}{\mathbf{w}}
\newcommand{\bU}{\mathbf{U}}
\newcommand{\bV}{\mathbf{V}}
\newcommand{\bF}{\mathbf{F}}
\newcommand{\bB}{\mathbf{B}}
\newcommand{\bk}{\mathbf{k}}
\newcommand{\Lop}{{\cal L}}
\newcommand{\Fop}{{\cal F}}
\newcommand{\Dofr}{{\cal D}(r)}
\newcommand{\Dt}{\Delta t}
\newcommand{\bbA}{\mathbf{A}}
\newcommand{\bbZ}{\mathbf{Z}}
\newcommand{\bbK}{\mathbf{K}}
\newcommand{\bbI}{\mathbf{I}}
\newcommand{\bbb}{\mathbf{b}}
\newcommand{\bbe}{\mathbf{e}}
\newcommand{\bbone}{\mathbf{1}}

\newcommand{\dx}{\Delta x}
\newcommand{\dt}{\Delta t}
\newcommand{\hfp}{\hat{f}_{j+\half}}
\newcommand{\hfn}{\hat{f}_{j-\half}}
\newcommand{\aik}{\alpha_{i,k}}
\newcommand{\bik}{\beta_{i,k}}
\newcommand{\lt}{\tilde{L}}

\newcommand{\hf}{\frac{1}{2}}
\newcommand{\fracStrut}{\rule[-1.0ex]{0pt}{3.1ex}}
\newcommand{\hfs}{\ensuremath{\frac{1}{2}}\fracStrut}
\newcommand{\scinot}[2]{\ensuremath{#1\times10^{#2}}}
\newcommand{\dee}{\mathrm{d}}
\newcommand{\dye}{\partial}
\newcommand{\diff}[2]{\frac{\dee #1}{\dee #2}}
\newcommand{\pdiff}[2]{\frac{\dye #1}{\dye #2}}
\newcommand{\Real}{\mathbb{R}}
\newcommand{\Complex}{\mathbb{C}}
\newcommand{\m}[1]{\mathbf{#1}}
\newcommand{\mA}{\m{A}}
\newcommand{\mI}{\m{I}}
\newcommand{\mK}{\m{K}}
\newcommand{\mL}{\m{L}}
\newcommand{\matalpha}{\boldsymbol{\upalpha}}
\newcommand{\matbeta}{\boldsymbol{\upbeta}}
\newcommand{\matmu}{\boldsymbol{\upmu}}
\newcommand{\matgamma}{\boldsymbol{\upgamma}}
\newcommand{\matlambda}{\boldsymbol{\uplambda}}
\renewcommand{\v}[1]{\boldsymbol{#1}}
\newcommand{\transpose}{^\mathrm{T}}
\newcommand{\bT}{\v{b}\transpose}
\newcommand{\vb}{\v{b}}
\newcommand{\vc}{\v{c}}
\newcommand{\ve}{\v{e}}
\newcommand{\vu}{\v{u}}
\newcommand{\vv}{\v{v}}
\newcommand{\vy}{\v{y}}
\newcommand{\Matlab}{{\sc Matlab}\xspace}
\newcommand{\code}[1]{\textsf{#1}}
\newcommand{\sspcoef}{\mathcal{C}}

\newcommand{\Fig}[1]{Figure~\ref{fig:#1}}
\newcommand{\cmean}{\hat{c}}
\newcommand{\rhomean}{\bar{\rho}}
\newcommand{\Kmean}{\hat{K}}
\newcommand{\Kinvmean}{\left<K^{-1}\right>}
\newcommand{\Zmean}{\left<Z\right>}



\section{Introduction\label{intro}}
Consider one-dimensional nonlinear wave propagation in a spatially
heterogeneous medium, described by the first order hyperbolic system
\begin{subequations} \label{nel_pde}
\begin{align}
\epsilon_t(x,t)-u_x(x,t) & = 0 \\
\rho(x)u(x,t)_t - \sigma(\epsilon(x,t),x)_x & = 0.
\end{align}
\end{subequations}
This system is a rather generic description of nonlinear waves in a 
Lagrangian frame, and arises in a variety
of contexts including elasticity, optics, and gas dynamics.
In the case of elasticity, $\epsilon,\sigma,\rho$, and $u$ are the
strain, stress, density, and velocity, respectively.

All the results presented in this work involve a simple periodic medium
composed of alternating homogeneous layers of materials A and B:
\begin{align} \label{LYmedium}
(\rho(x),K(x)) & = \left\{ \begin{array}{ll}
 (\rho_A,K_A) & \text{if } j< x < (j+1/2)
   \mbox{ for some integer j}, \\
 (\rho_B,K_B) & \mbox{otherwise.} \end{array}\right.
\end{align}
with nonlinear stress-strain relation
\be \label{expstress}
\sigma(\epsilon,x) = \exp(K(x)\epsilon) - 1,
\ee
Further computational experiments, to be reported elsewhere, suggest that 
the qualitative nature of our findings is typical for propagation in
more general periodic materials with quite general nonlinearities.


\subsection{Some suggestive numerical experiments}
\begin{figure}
\centerline{
\includegraphics[width=2.5in]{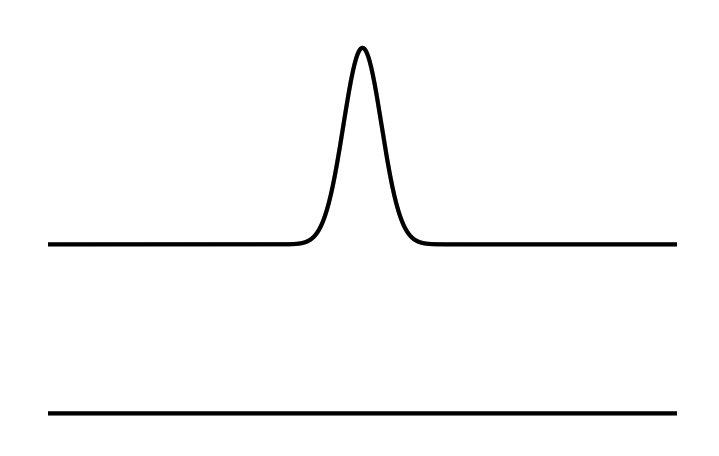}}
\caption{Initial condition for the three experiments.  In this and subsequent plots,
the upper plot is stress and the lower plot is velocity.\label{fig:ic}}
\end{figure}

\begin{figure}
\centerline{
\subfigure[Solution at $t=60$\label{fig:experiment1_middle}]{\includegraphics[width=3.0in]{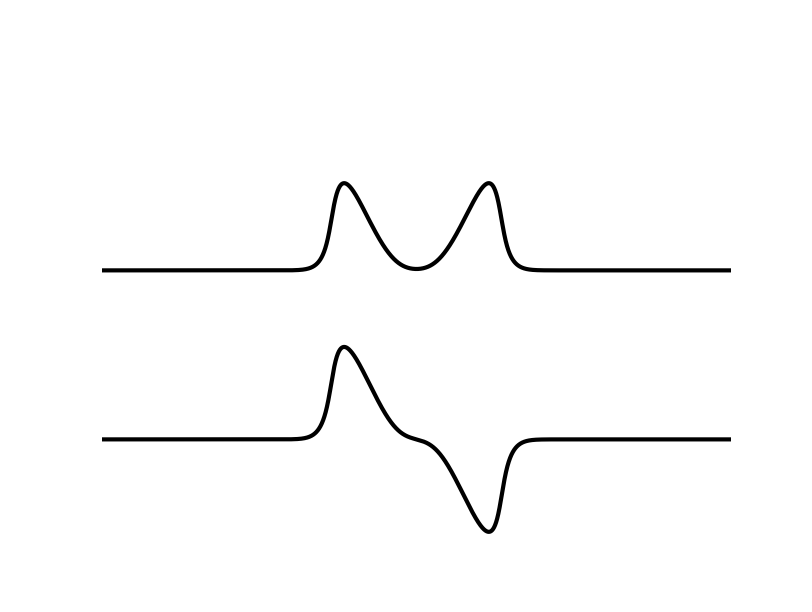}}
\subfigure[Solution at $t=120$\label{fig:experiment1_end}]{\includegraphics[width=3.0in]{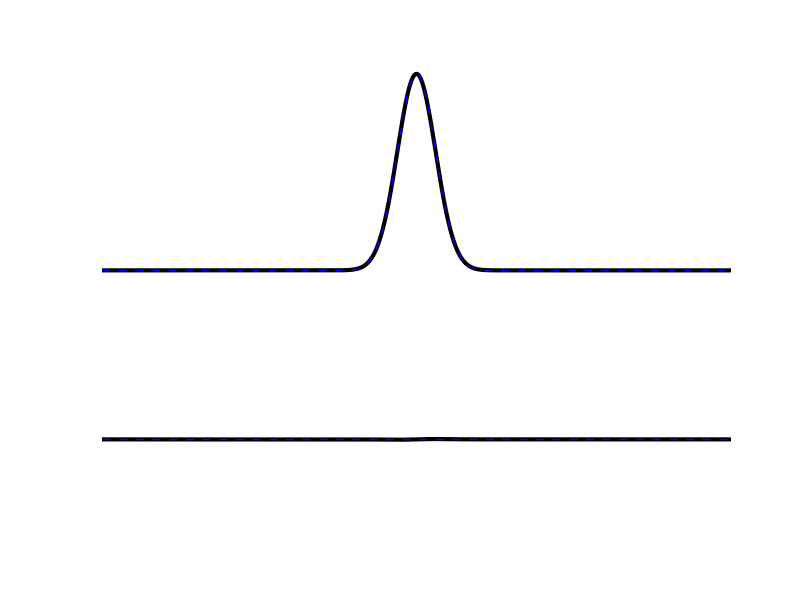}}}
\caption{Stress (upper plot) and velocity (lower plot) snapshots of experiment 1, 
in which no shocks form.  The final
solution is identical to the initial condition.\label{fig:experiment1}}
\end{figure}

To motivate the topic of this paper, we present the following simple experiments.
Consider the nonlinear wave equation
\eqref{nel_pde} with initial velocity zero and Gaussian initial stress, 
shown in Figure \ref{fig:ic}.
In the first experiment, we consider a homogeneous medium with $\rho(x)=K(x)=1$ 
and the solution is evolved for a short time. 
As shown in Figure \ref{fig:experiment1_middle}, the initial 
hump evolves into a left-going
and a right-going pulse.  At the time shown, the sign of $u$ is negated,
and then the solution is evolved
again for the same length of time.
The final solution is identical to the initial condition, as shown in Figure
\ref{fig:experiment1_end}.  The second half of the
experiment is, of course, identical to the first half, but in reverse.  

\begin{figure}
\centerline{
\subfigure[Solution at $t=250$\label{fig:experiment2_middle}]{\includegraphics[width=3.0in]{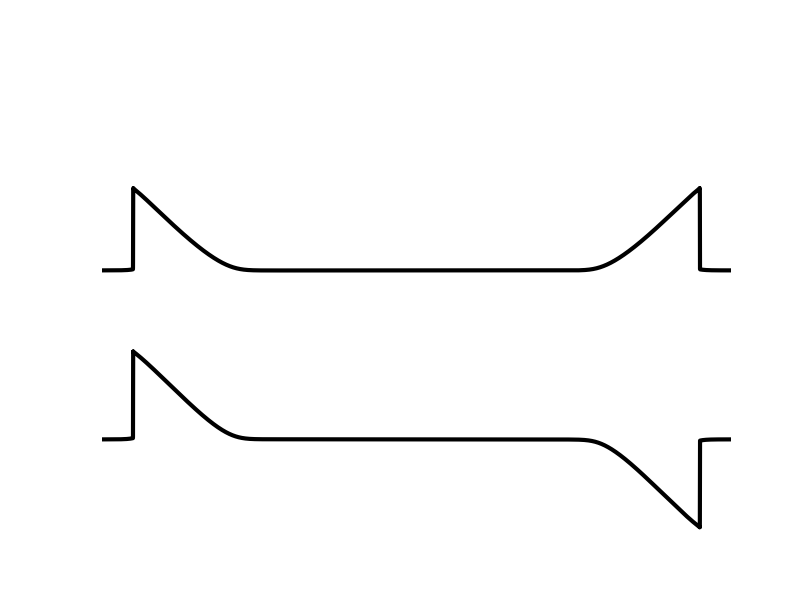}}
\subfigure[Solid line is solution at $t=500$; dotted line is initial condition\label{fig:experiment2_end}]{\includegraphics[width=3.0in]{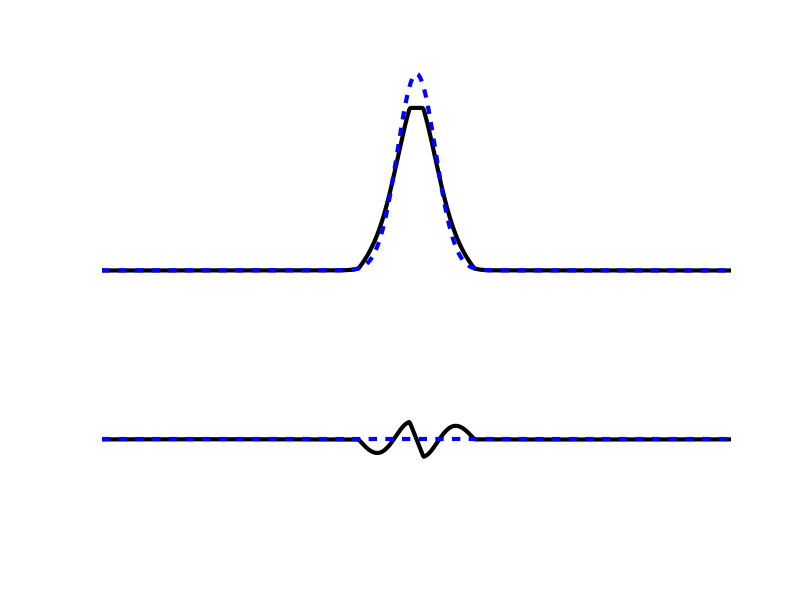}}}
\caption{Stress and velocity snapshots of experiment 2, in which shocks form.\label{fig:experiment2}}
\end{figure}

In the second experiment, again $\rho(x)=K(x)=1$ but now the solution is evolved
to a much later time.  
As shown in Figure \ref{fig:experiment2_middle}, by this time the left- and 
right-going
pulses have developed shocks.  Again the velocity is reversed and the solution is
evolved for the same length of time.  
In this case the final solution, shown in Figure \ref{fig:experiment2_end} 
is quite different from the initial condition.

\begin{figure}
\centerline{
\subfigure[Solution at $t=250$]{\includegraphics[width=3.0in]{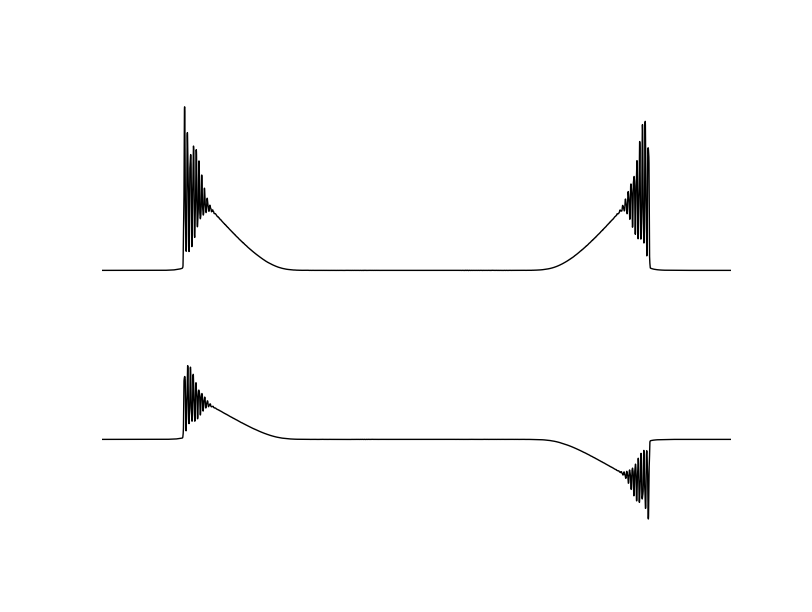}}
\subfigure[Solution at $t=500$]{\includegraphics[width=3.0in]{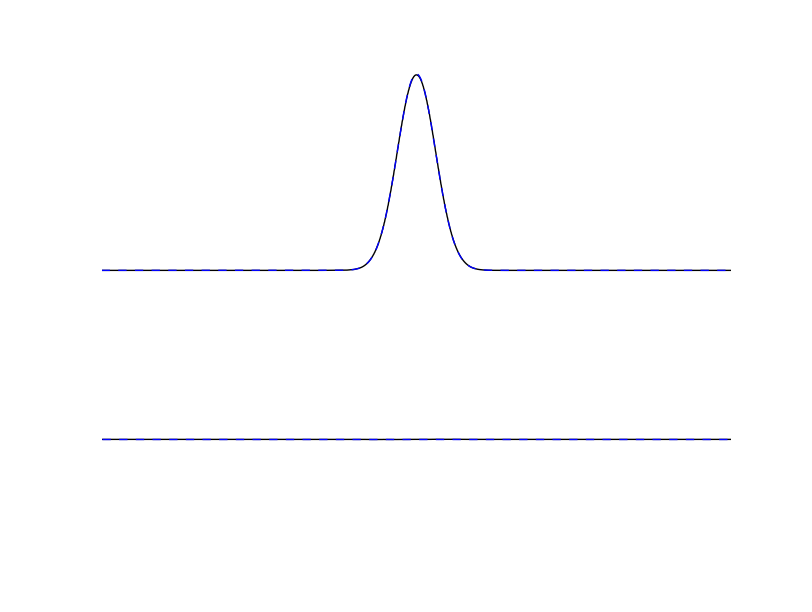}}}
\caption{Stress and velocity snapshots of experiment 3, in a periodic medium.  The final solution is identical to the initial condition.\label{fig:experiment3}}
\end{figure}

In the third experiment, the medium is taken to be periodic and piecewise
constant, composed of alternating homogeneous layers of materials A and B,
as described by \eqref{LYmedium}, with $\rho_A=K_A=1$ and $\rho_B=K_B=4$.
The solution is evolved to the same time as in experiment two above. 
This time, highly oscillatory fronts develop, as shown in Figure \ref{fig:experiment3}.
Once again the velocity is reversed and 
the solution evolved for the same length of time. 
In this case the final solution appears identical to the initial condition.

\subsection{First-order hyperbolic systems theory}
The results of the first two experiments are well understood within the existing
mathematical theory of hyperbolic conservation laws.  The system \eqref{nel_pde}
is time-reversible as long as the solution remains smooth.
In the first experiment, since shocks do not form,
the solution satisfies the nonlinear wave equation \eqref{nel_pde} 
in a strong sense and is time-reversible.
In the second experiment, characteristics meet and shocks form, leading to
a loss of information.  In fact, whereas the initial condition we have used
is the only one
that leads to the solution shown in Figure \ref{fig:experiment1_middle}
at the given time, there are infinitely many initial conditions that lead
to the solution shown in Figure \ref{fig:experiment2_middle}; the 
curves in Figure \ref{fig:experiment2_end} are two of them.
In general, such irreversible behavior is
expected whenever the solution is evolved past the time of shock formation.
Thus, from the point of view of the classical theory for nonlinear hyperbolic
systems, the long-time reversibility of waves in the periodic
medium observed in the third experiment seems remarkable.
Similar waves in a homogeneous medium composed of material A
or material B would not be reversible at this late time. 


\subsection{Dispersive nonlinear wave theory}
It was shown in \cite{santosa1991} that for linear waves whose wavelength is
long relative to the period of the medium, the leading order
effect of material periodicity is an effective dispersion.
Correspondingly, dispersive wave equations are often introduced to model
the effect of periodic microstructure \cite{Rubin2009}.
Indeed, a system of dispersive effective equations for the elasticity system
\eqref{nel_pde} with periodic coefficients was derived in \cite{leveque2003} 
and shown to agree with results from direct simulation.
Dispersive nonlinear wave equations lead to the appearance of so-called
{\em dispersive shock waves}, consisting of a high-amplitude oscillatory
front followed by a more slowly varying tail \cite{El2005}.
The waves in \Fig{experiment3} strongly resemble these dispersive shocks.
In such systems, the dispersive term(s) in the equation
can often be shown to regularize the solution, preventing the appearance of 
discontinuities.  Hence time-reversibility is typically a property of such
dispersive nonlinear systems.

However, the effective dispersive equations of \cite{leveque2003}, like
many dispersive continuum models, rely on an assumption that the wavelength of
the solution is large relative to the period of the medium.  Since nonlinearity
leads to the appearance of high frequencies in the solution, it seems at
least possible that this model will break down.  One may ask, then,
whether true shocks (discontinuities) may indeed appear.

The answer to this question turns out to be quite interesting.  As suggested
already by the third experiment above,
it appears that shocks do not form, even after very long times
so long as the amplitude of the initial conditions is not too large 
relative to the effective dispersion induced by material periodicity.
Furthermore, initial shocks with small amplitude appear to be unstable
and vanish after a short time.
For larger-amplitude solutions, (or weaker effective material dispersion) 
shock discontinuities appear and persist.
Empirically, we find an approximate condition discriminating between data 
that will or will not lead to shocks, in the form of a characteristic condition 
that can be seen as a 
generalization of the well-known Lax entropy condition for shock admissibility.

In the remainder of this section, we describe briefly the numerical
methods used in this work.  In Section \ref{measures},
we discuss the problem of detecting shock formation and propose two
robust computational approaches.  Along the way, we make some 
observations about limiters used in high-resolution shock-capturing methods.  In Section 
\ref{criterion}, we hypothesize
a condition for shock formation in solutions of the nonlinear wave equation
\eqref{nel_pde} in the presence of a periodic medium and
conduct some further numerical tests for layered media that support the
proposed condition.  In Section \ref{discussion}, 
we discuss the significance of the results and possible generalizations.

The results presented here can be understood even better when accompanied
by animations of the wave behavior described.  These are available online,
along with all code for reproducing the computational results described,
at \url{http://bitbucket.org/ketch/layeredmediashocks/src}.
The reader is highly encouraged to view the animations and experiment
with the simulations.

\subsection{Numerical discretization\label{discretization}}
Before continuing, we briefly describe the numerical methods used for the
computations presented in this work.  The methods employed are both
finite volume Godunov-type
high-resolution methods, which employ Riemann solvers and nonlinear limiters
to obtain good resolution of shocks or steep gradients without spurious
oscillations.  

The first method used is that implemented in Clawpack \cite{clawpack45}
and described in \cite{leveque1997}.  Briefly, this is a second-order TVD
high-resolution scheme based on Lax-Wendroff discretization with limiters.

The second method used is that implemented in the SharpClaw software package
\cite{sharpclaw} and described in \cite{ketcheson2006,ketcheson2011}.  
This involves a method-of-lines
discretization approach, using WENO reconstruction in space and high order
Runge-Kutta time integration.  In all experiments with SharpClaw we use the
fourth-order SSP Runge-Kutta scheme of \cite{ketcheson2008} and fifth-order 
WENO reconstruction.

\section{Computational Measures of Shock Formation\label{measures}}
  It is clear, at least in the case of a piecewise homogeneous medium, 
that at least some kinds of smooth initial data will lead to shocks,
regardless of the material parameters.  For instance,
if the initial data includes sufficiently large gradients, certain
characteristics will intersect before they reach a material interface.
In fact, for any fixed medium it is possible to construct initial data of
arbitrarily small amplitude for which a shock forms in the solution.  To
construct such data, choose a point $x_0$ so that $\sigma(\epsilon,x)$ is
continuous in $x$ in an open neighborhood about $x_0$ and take data
\begin{subequations} \label{ic1}
\begin{align}
\epsilon(x,-\tau) &= \begin{cases} \epsilon_0 &\text{for}~ x<x_0\\
0 &\text{for}~ x\geq x_0, \end{cases}\\
\noalign{\vskip 3pt}
u(x,-\tau) &= \begin{cases} -\epsilon_0\sqrt{\frac{\sigma'(\epsilon_0)}{\rho(x_0)}}
 &\text{for}~ x<x_0\\
0 &\text{for}~ x\geq x_0 \end{cases}
\end{align}
\end{subequations}
at some very small negative time $-\tau$ with $0<\tau\ll 1$.
This data contains a jump discontinuity that spreads out as a 1-rarefaction
wave, so that solving up to time $t=0$ gives smooth functions
$\epsilon(x,0)$ and $u(x,0)$.  Now negate $u(x,0)$ and consider the
resulting functions as data at $t=0$.  By time-reversibility, over the time
$0\leq t\leq \tau$ the solution sharpens back into the discontinuity we
started with, corresponding to shock formation.

However, the fact that we can construct data resulting in shock waves 
does not preclude the possibility that smooth solutions
exist for all time for some restricted set of initial data.



Detecting the formation of shocks in computed solutions is challenging,
since the computation produces only a finite number of values (cell-averages
in the case of the finite volume methods used here) and in general
it is impossible to determine if the solution is smooth based on these
values.  In practice, one can only expect to obtain an upper-bound on the
magnitude of possible discontinuities.
In \cite{simpson2010}, visual detection of shocks in spectral solutions
was conducted by looking for highly oscillatory regions near steep gradients
(evidence of the Gibbs phenomenon).
Since the solutions we are interested in contain highly oscillatory regions
and steep smooth regions, inspection of the solution may not be a reliable
way to judge whether shocks have formed.  We propose a more robust and
quantitative approach based on two
signatures of shock formation.  The first, mentioned already in the 
introduction, is the {\em loss of time-reversibility}.  The second is 
{\em entropy decay}.  
We now describe the design of experiments that can detect these
signatures.

  \subsection{Time Reversal}
    Solutions of hyperbolic systems are time-reversible as long as shocks
do not form.  This can be used as a computational tool to probe
regularization, following the approach used already in the experiments
of Section \ref{intro}:
\begin{enumerate}
  \item Begin with smooth initial data $q_0$.
  \item Numerically solve \eqref{nel_pde} up to time $T$ to obtain a solution $q_T$.
  \item Reverse the sign of the velocity $u$, to obtain a new solution $q_0^*$.
  \item Numerically solve \eqref{nel_pde} from time $T$ to time $2T$ 
        with initial condition $q_0^*$ to obtain a solution $q_T^*$.
  \item If no shocks formed, the solution $q_T^*$ should match $q_0$, up to
          numerical errors.
\end{enumerate}

\begin{figure}
\centerline{
\subfigure[Initial pulse]{\includegraphics[width=2.5in]{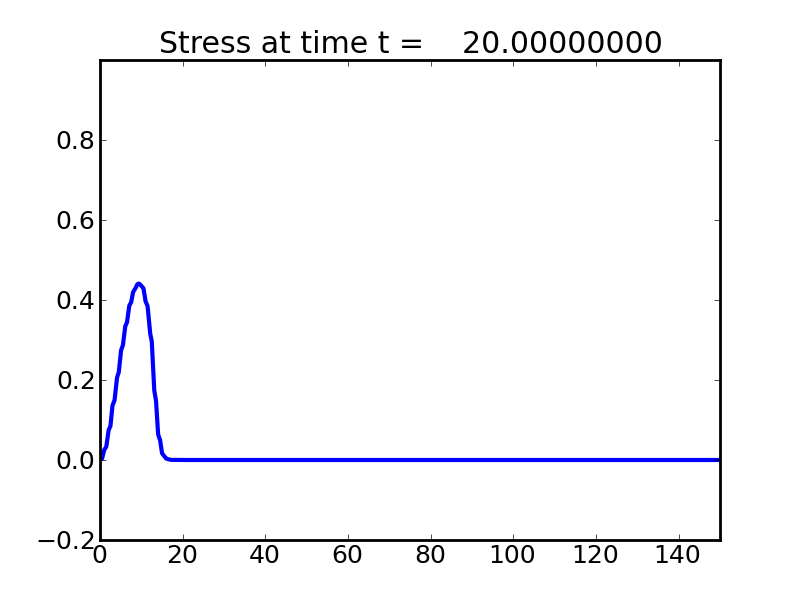}}
\subfigure[Separation into solitary waves]{\includegraphics[width=2.5in]{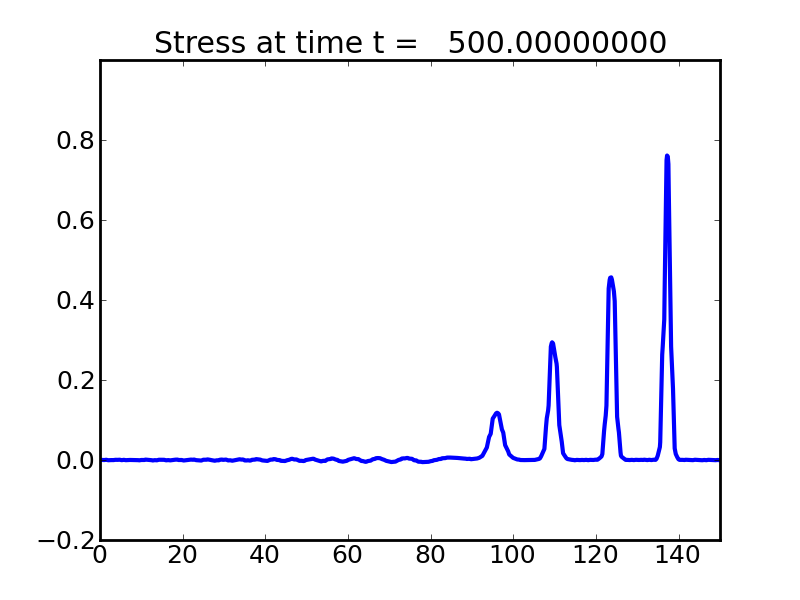}}}
\caption{Evolution of a single pulse into a solitary wave train. \label{fig:stego}}
\end{figure}

As a demonstration, consider the problem from \cite{leveque2003}, 
which we will call the LY problem
henceforth.  This problem is defined as follows.
We consider the exponential stress relation \eqref{expstress}, and the layered 
medium defined by \eqref{LYmedium} with
$\rho_A=K_A=1 \mbox{  and  } \rho_B=K_B=4$.
An initial pulse is generated by motion of the left boundary:
\begin{align*}
u(0,t) & = \begin{cases} 
  -0.1(1+\cos(\pi(t-10)/10)) & \mbox{for } 0\le t\le 20 \\
  0 & \mbox{for } t>20. \end{cases}
\end{align*}
After a short time, periodic boundary conditions are imposed in
order to observe the long-time behavior of the pulse without using
an excessively large computational domain.
The initial half-cosine pulse evolves into a train
of solitary waves, as shown in Figure \ref{fig:stego}.

\begin{figure}
\centerline{
\subfigure[SharpClaw\label{fig:stego_tr_sc}]{\includegraphics[width=3in]{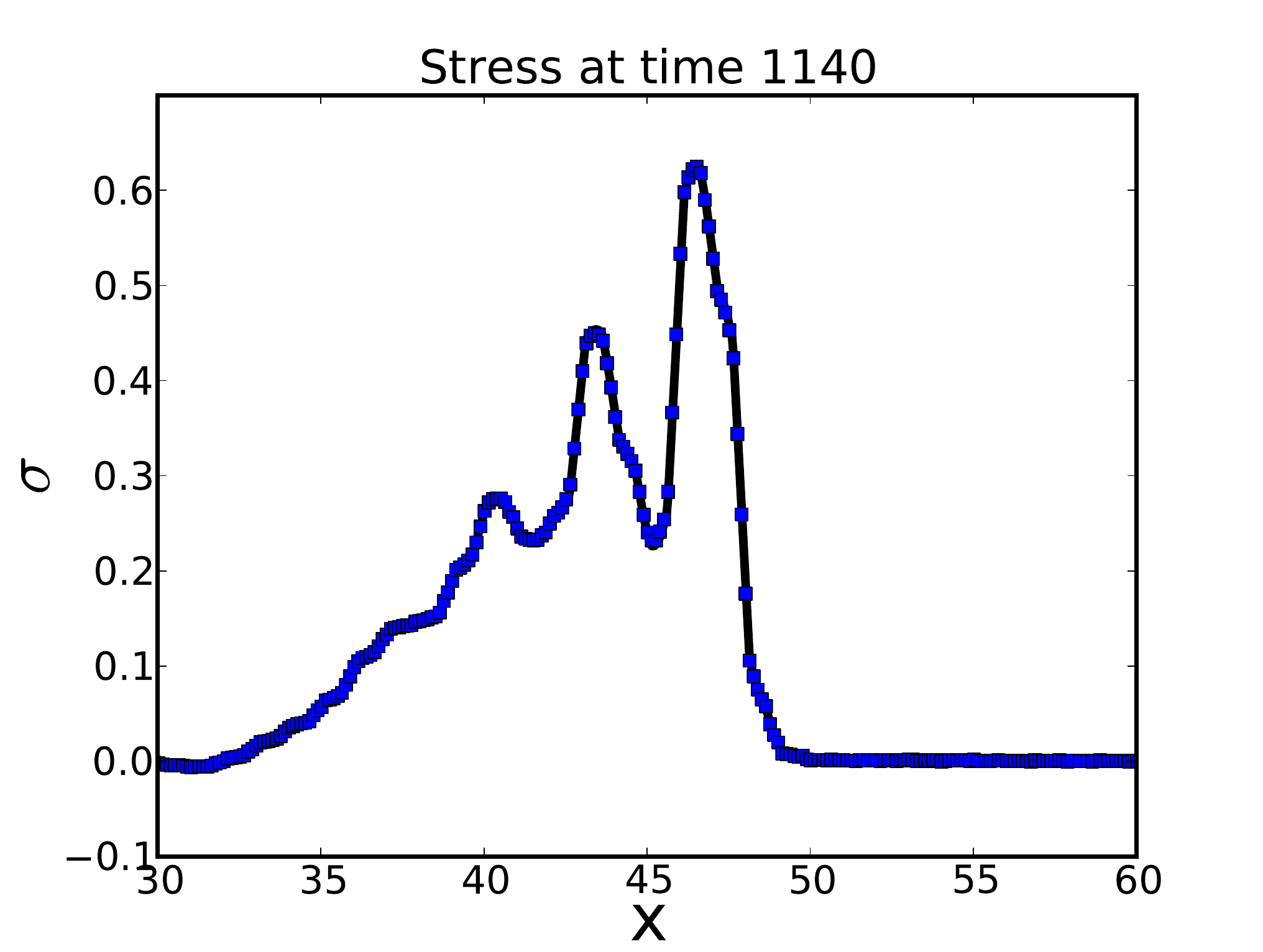}}
\subfigure[Clawpack\label{fig:stego_tr_cp}]{\includegraphics[width=3in]{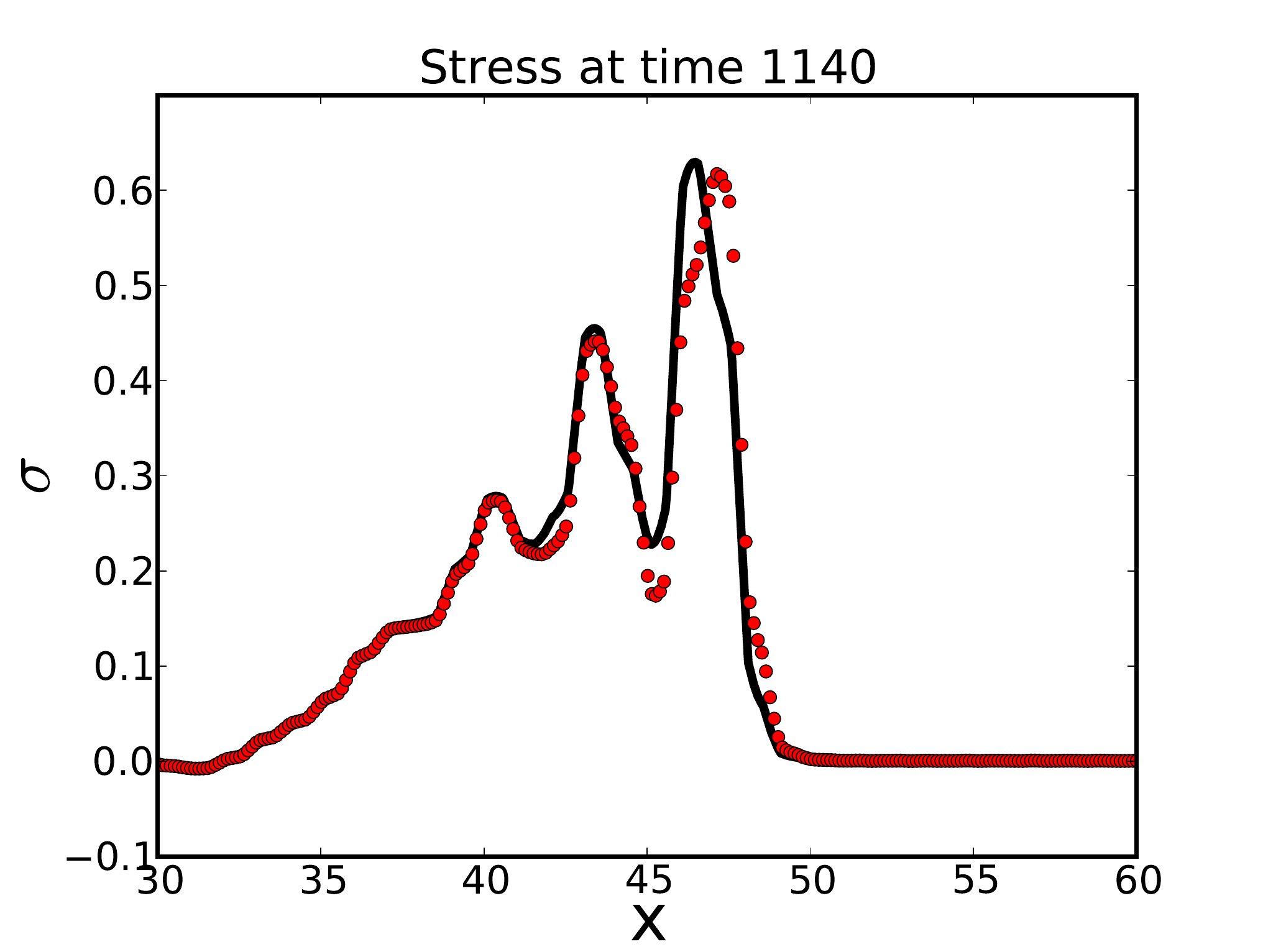}}}
\caption[Comparison of forward solution and time-reversed solution 
stegotons.]{Comparison of forward solution (black line) and 
time-reversed solution (symbols).\label{fig:stego_tr}}
\end{figure}

We use the solution to the LY problem at $t=40$ as
initial data, solving up to $T=600$, reversing the velocity, and solving again
up to time $1160$.  
The solutions at these two times,
plotted in \Fig{stego_tr_cp}, appear nearly identical.
Indeed, the maximum pointwise difference of the
initial and final velocities,
\be \label{eq:discrepancy}
E=\|u(x,1160)-u(x,40)\|_\infty,
\ee
which we refer to as the {\em discrepancy}, is quite small.
In \Fig{stego_tr_sc} we plot the 
solution obtained using the SharpClaw software \cite{ketcheson2006} on a grid with 24 cells
per layer.  The $t=1160$ solution (blue squares) is in excellent agreement
with the $t=40$ solution (black line).  
For comparison we also show a solution obtained using Clawpack
on the same grid (24 cells per
layer), in \Fig{stego_tr_cp}.  Clawpack uses second order accurate methods
with limiters, and gives a less accurate 
(but also less computationally expensive) solution on the same grid.

In general the maximum error $E$ is small; We would like to verify 
that the difference between 
initial and final solutions is purely due to numerical errors, and not
due to the formation of (time-irreversible) shocks.  We thus consider
the rate at which $E$ decreases as the grid is refined, using the two 
different numerical methods.
Table \ref{tbl:trconv} lists 
the discrepancy $E$ obtained for a range of grids with both Clawpack and SharpClaw.
By using finer grids, both solutions appears to converge to the early time solution.
The convergence rate of the SharpClaw scheme fluctuates considerably, and
ongoing work is aimed at understanding this.

\begin{table} \centering
\begin{tabular}{r|cc|cc} \hline
& \multicolumn{2}{c}{Clawpack} & \multicolumn{2}{c}{SharpClaw} \\
N & $E$ & Rate & $E$ & Rate \\ \hline
12 & $4.36 \times 10^{-1}$ & -    &  $5.89 \times 10^{-1}$ & -    \\
24 & $8.15 \times 10^{-2}$ & 2.42 &  $8.14 \times 10^{-3}$ & 6.18 \\
48 & $1.54 \times 10^{-2}$ & 2.40 &  $9.81 \times 10^{-4}$ & 3.05 \\
96 & $3.29 \times 10^{-3}$ & 2.23 &  $3.72 \times 10^{-4}$ & 1.40 \\ 
192& $7.41 \times 10^{-4}$ & 2.15 &  $7.86 \times 10^{-5}$ & 2.24 \\ 
\hline
\end{tabular}
\caption{Maximum pointwise discrepancy ($E$ defined in \ref{eq:discrepancy}) for 
time-reversal test using Clawpack and Sharpclaw.
The quantity $N$ is the number of computational cells per layer of the
medium.\label{tbl:trconv}}
\end{table}

\begin{figure} \centering
\subfigure[24 cells per layer]{\includegraphics[width=3in]{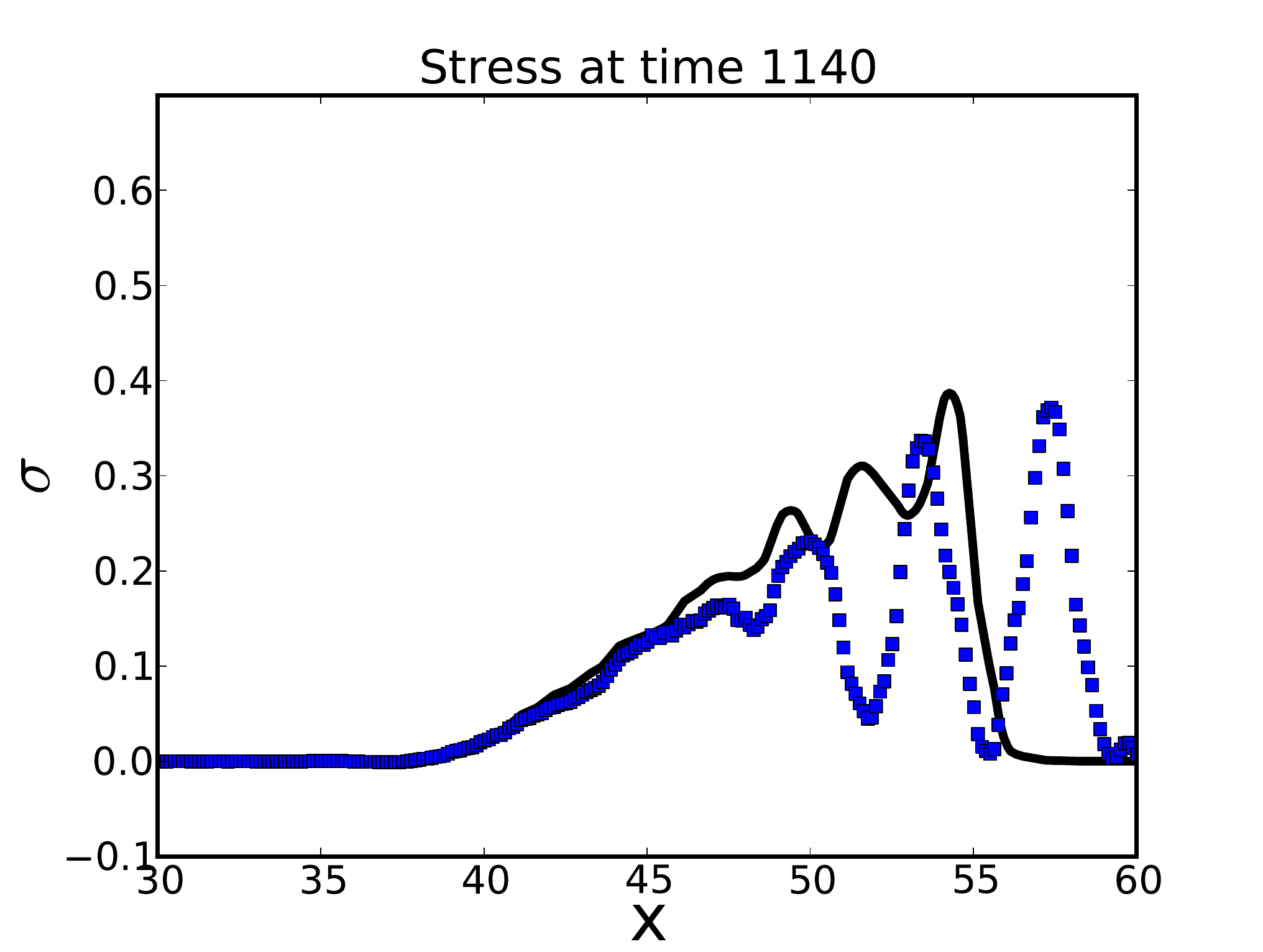}}
\subfigure[48 cells per layer]{\includegraphics[width=3in]{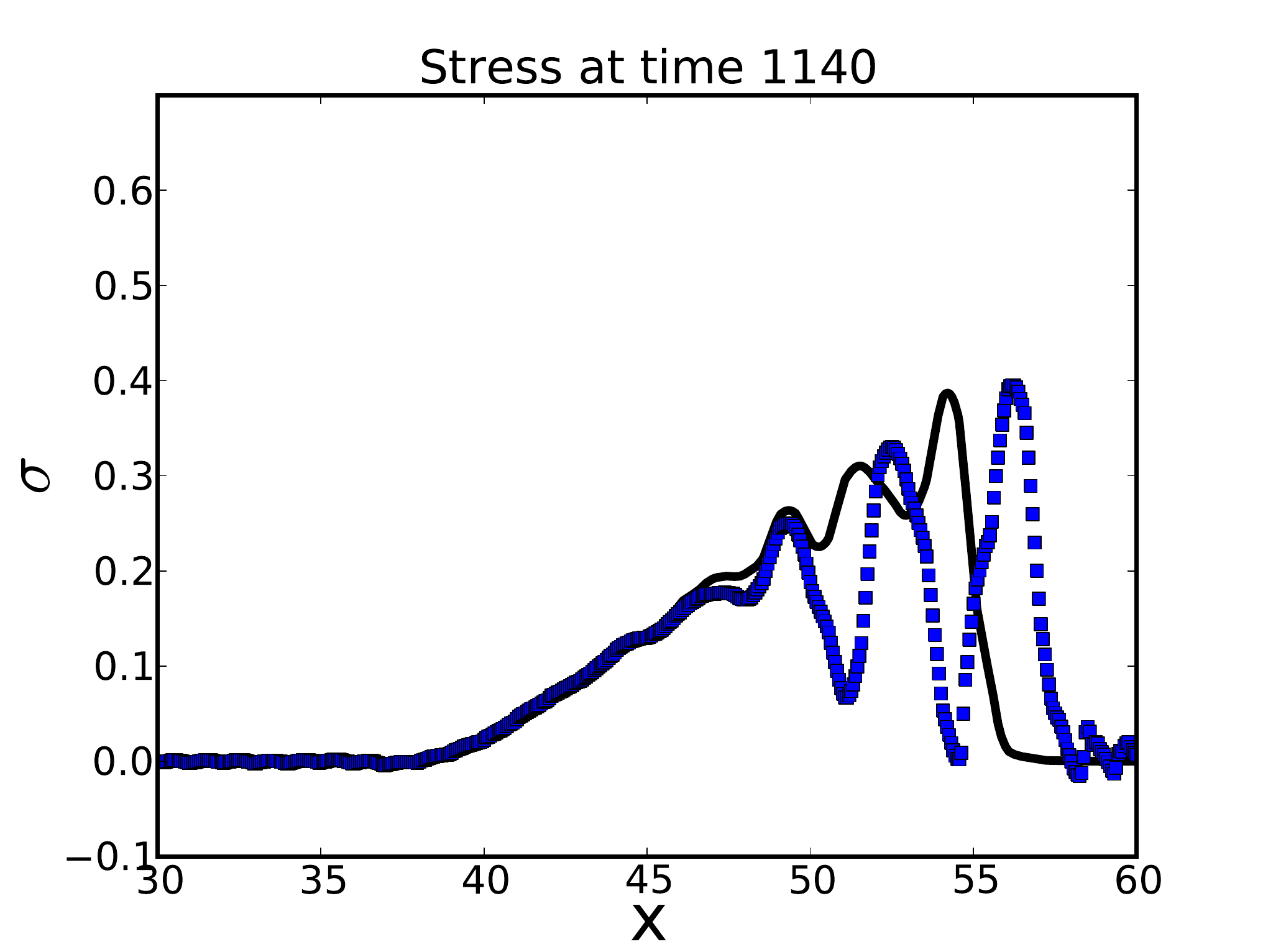}}
\caption{Comparison of forward solution (solid line) and 
time-reversed solution (dotted line) obtained with SharpClaw 
using 24 and 48 cells per layer, for an LY medium
with lower impedance contrast $Z_B=2$.
The solution seems not to be time-reversible.
\label{fig:stego_ntr}}
\end{figure}

Next we conduct the same test but with a less-strongly varying medium,
taking $\rho_B=K_B=Z_B=2$.
\Fig{stego_ntr} shows the results obtained with SharpClaw using 24 and
48 cells per layer.  Observe the large 
difference between the initial and final solutions.
This seems to indicate that shock formation has occurred in 
this case, leading to a loss of time-reversibility.

  \subsection{Entropy Evolution}
    Another way to detect the formation of shocks computationally is by
measuring entropy.  An entropy function for \eqref{nel_pde} is the total energy:
\begin{align}
  \eta(u,\epsilon,x) = \frac{1}{2}\rho(x)u^2 + \int_0^\epsilon \sigma(s,x)ds.
\end{align}
It is straightforward to see that $\eta$ is conserved for smooth solutions:
\begin{align*}
  \frac{d}{dt} \int_{-\infty}^\infty \eta dx & = \int_{-\infty}^\infty \eta_t dx \\
    & = \int_{-\infty}^\infty \left(\rho(x) u u_t + \frac{d}{dt}\int_0^{\epsilon(x,t)} \sigma(s,x) ds \right) dx \\
    & = \int_{-\infty}^\infty \left(\rho u u_t + \sigma(\epsilon,x)\epsilon_t\right) dx \\
    & = \int_{-\infty}^\infty \left(u\sigma_x + \sigma u_x\right) dx = \int_{-\infty}^\infty (\sigma u)_x dx = 0.
\end{align*}
However, when shocks form the entropy of the physically correct
vanishing-viscosity solution will decrease.  Since our numerical
methods are designed to compute the vanishing-viscosity solution,
the numerical entropy will also tend to decrease when shocks form.
In this section we study the evolution of entropy in computational solutions.

\begin{figure}
\centerline{
\includegraphics[width=4in]{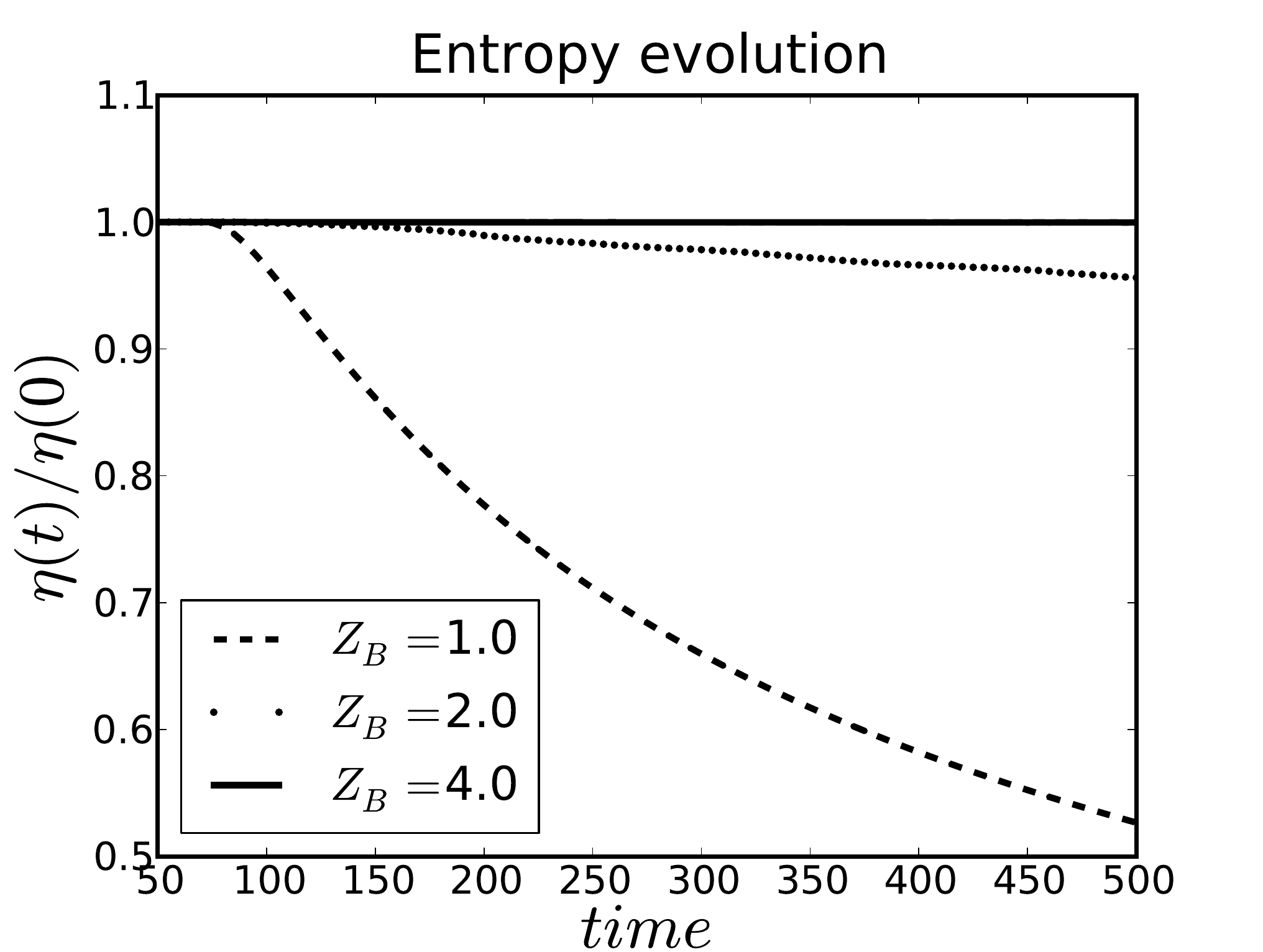}}
\caption{Entropy evolution in time for media with various impedance
ratios.  In this test, the density and bulk modulus of layer B were 
varied simultaneously and are equal to $Z_B$ in each case.\label{fig:entropy_1}}
\end{figure}

In \Fig{entropy_1}, we show the evolution of entropy over time for 
an initial gaussian stress perturbation
in a medium with $\rho_A=K_A=1$ and $\rho_B=K_B=Z_B$, for
varying values of $Z_B$.  In each case, the entropy is normalized to be
unity at time zero.  The case $Z_B=1$ corresponds to a homogeneous
medium, and the entropy decays rapidly once a shock forms.  
For $Z_B=2$, the entropy evolution indicates that shock formation is
delayed and the resulting shocks are weaker.
For $Z_B=4$, the entropy is nearly constant over the duration
of the simulation.

\begin{figure}
\centerline{
\includegraphics[width=4in]{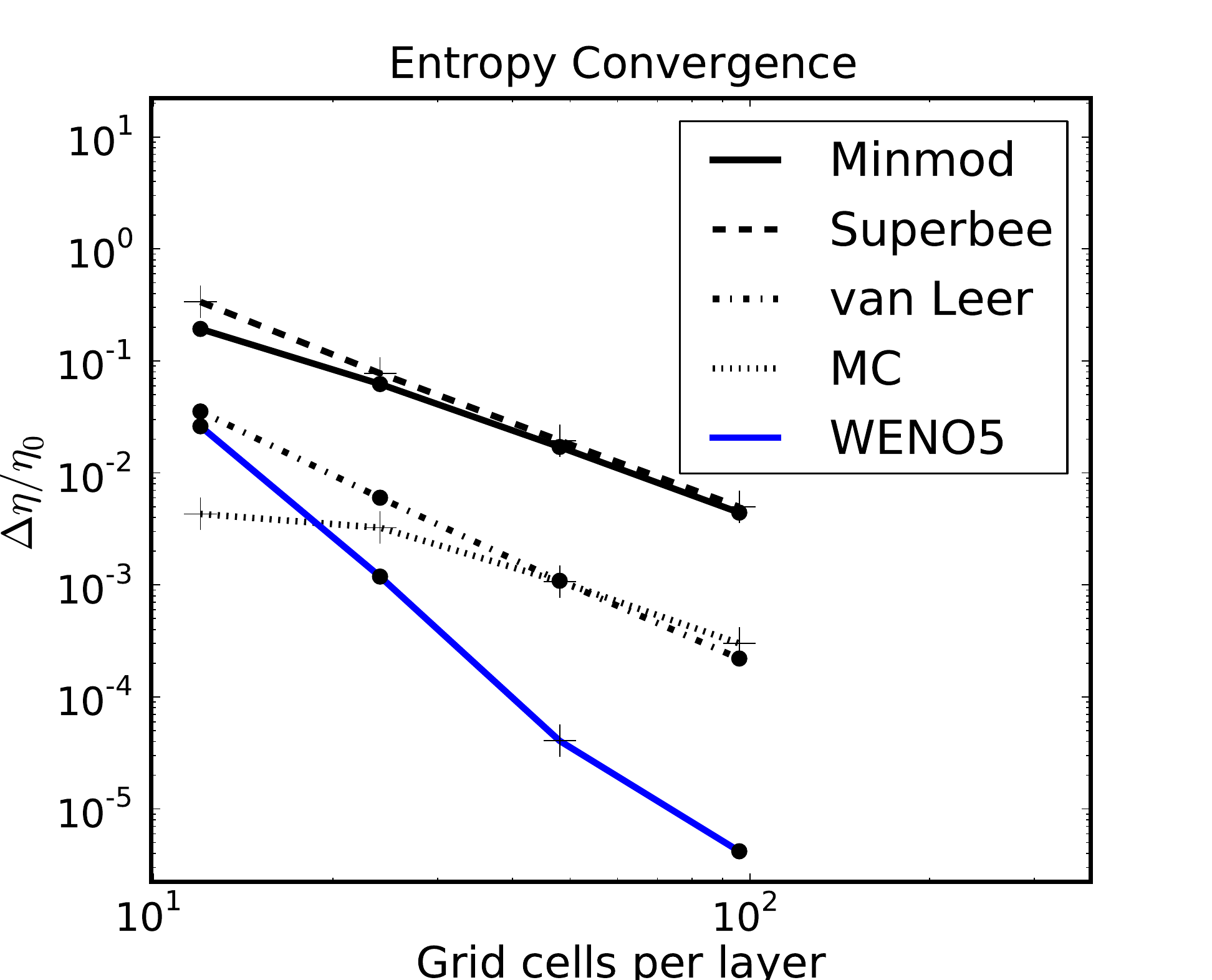}}
\caption{Entropy change up to $T=500$ versus number of grid cells per
medium layer for the four limiters implemented in Clawpack.  Here $Z_B=4$.
Plus signs ('+') indicate cases for which the entropy has increased, while
filled circles
indicate cases for which the entropy has decreased.\label{fig:entropy_lims}}
\end{figure}

Numerical dissipation can also lead to a loss of entropy in the 
computational solution.  Using a first-order method, the entropy evolution
is entirely dominated by this effect.  For the higher order numerical
methods employed in this paper, we have found that entropy may 
numerically increase or decrease
depending on how agressive the limiter is.  Overcompressive numerical 
limiters can lead to (unphysical) entropy production.  However, all
numerical effects on entropy evolution appear to decrease rapidly
with grid resolution, as shown in \Fig{entropy_lims}, which indicates
the relative entropy loss versus grid resolution at $T=500$ for $Z_B=4$.
Among the limiters implemented in Clawpack, the smallest change 
in entropy is generally produced by the van Leer limiter (for sufficiently
fine grids).
The van Leer limiter is used in all subsequent Clawpack simulations 
in this paper.  Using SharpClaw with fifth-order WENO reconstruction, 
we find that the entropy errors are much smaller than for any of the 
Clawpack limiters, especially on the finer grids.

We remark that this test might be taken as a measure of the 
 dissipativity of a given limiter.  For instance, note that
the results in \Fig{entropy_lims} for Minmod and Superbee suggest that
they are dissipative and over-compressive, respectively.
It is interesting to note that, by this measure, the MC limiter
is somewhat over-compressive, while the van Leer limiter is somewhat
dissipative. Fifth-order WENO is neutrally-compressive.  That is, the
entropy fluctuates slightly around the initial value, but is roughly constant
in time if averaged over short intervals.

Due to numerical dissipation (or compression), it is impossible 
to completely rule out physical loss of entropy.  Instead, we can bound the 
possible loss of entropy through use of very fine spatial resolution.
For sufficiently fine grids, we have noticed that the small
numerical entropy fluctuations are nearly time-reversible.  That is,
if we run the time-reversibility experiment of the last section and
consider the entropy at corresponding early and late times, the 
difference is much smaller than the (already small) difference between
the entropy at either time and the initial entropy.  This allows us
to probe the degree to which entropy is
conserved for different impedance contrasts.  

\begin{figure}
\centerline{
\includegraphics[width=4in]{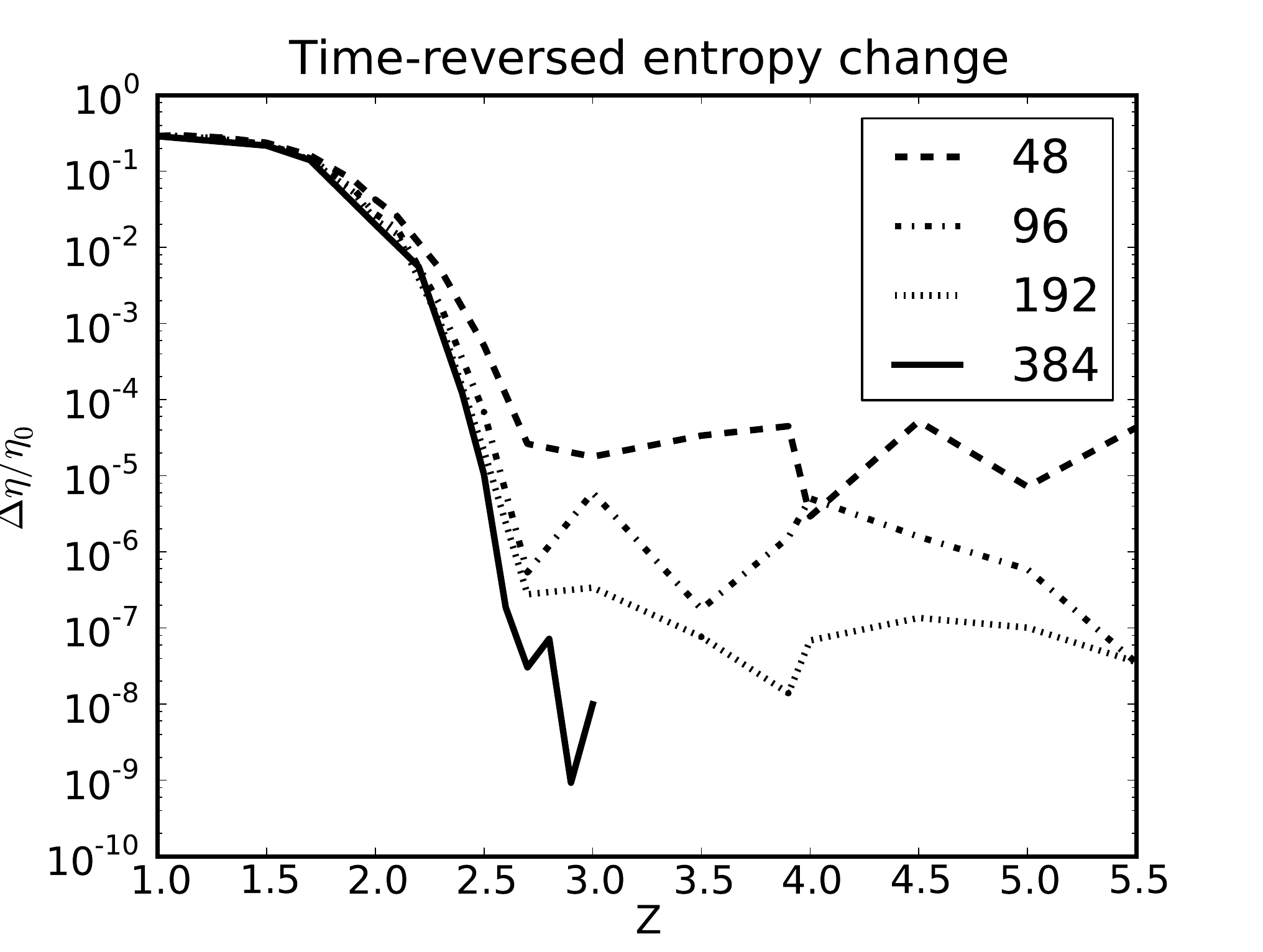}}
\caption{Change in entropy for LY media with increasing impedance contrast.
The different plots correspond to differing spatial resolutions; the
number of cells per material layer is indicated in the legend.\label{fig:trent}}
\end{figure}

In Figure \ref{fig:trent}, we plot the difference in entropy 
values at times $t=50$ and $t=450$ (using a final time of $T=500$), for 
SharpClaw solutions on fine grids.
We observe that the change in entropy generally 
gets smaller as the impedance contrast increases and as the resolution
of the simulation increases.  For a given grid, 
the entropy change apparently diminishes until it reaches the level of numerical
errors.  For impedance contrast greater than
2.5, the entropy change on the finest grid is less than $10^{-7}$.
Similar results are obtained with Clawpack, although much finer
resolution is required to resolve the small losses of entropy \cite{alghamdipetclaw}.

%
%

\section{A Condition for Shock Formation in Heterogeneous Media\label{criterion}}
    In this section we hypothesize and test a condition for shock formation.
This condition is empirically motivated, but leads to quantitatively accurate
predictions and can be motivated by simple arguments.

We will make use of the arithmetic and harmonic averaging operators, 
denoted by a bar and a hat, respectively:
\begin{align*}
\bar{f} & = \int_0^1 f(x) dx & \hat{f} & = \left(\int_0^1 \frac{1}{f(x)} dx\right)^{-1}.
\end{align*}

\subsection{Effective wave speeds}
In a homogeneous medium, small-amplitude perturbations travel at the
characteristic speed
\begin{align*}
c(\sigma_0) = \sqrt{\frac{\sigma'_0(\epsilon)}{\rho}} \ \ \ 
\mbox{ where } \sigma'_0(\epsilon)=\left.\sigma'(\epsilon)\right|_{\sigma=\sigma_0},
\end{align*}
where $\sigma_0$ is the stress in the ambient state.
In particular, in the linear case
these velocities are equal to the sound speed:
\begin{align} \label{hom_sound_speed} 
\sigma(\epsilon) & = K \epsilon  \implies
c =  \sqrt{K/\rho}. \end{align}


Now let us consider a linear periodic medium.  A pulse in such a medium
will spread out over time due to reflections.
The fastest moving part is that which is unreflected; this part travels at
the harmonic average of the sound speed:
\begin{align} \label{cmax}
\hat{c} = \left(\int_0^1{\sqrt{\frac{\rho(x)}{K(x)}} dx}\right)^{-1}.
\end{align}
Meanwhile, the main part of the signal (for a long-wavelength pulse) travels
at the effective sound speed given by using the average of the density and the
harmonic average of the bulk modulus \cite{santosa1991}:
\begin{align} \label{ceff}
c_\textup{eff} = \sqrt{\frac{\Kmean}{\rhomean}}
\end{align}
The effective velocity \eqref{ceff} is smaller than the maximum bulk
velocity \eqref{cmax},
due to the macroscopic slowing effect of repeated reflections.

By similar reasoning, small-amplitude perturbations traveling
in a nonlinear periodic medium with ambient stress $\sigma_0$
can be shown to travel at an effective velocity
\begin{align*}
c_\textup{eff}(\sigma_0) & = \pm\sqrt{\frac{\widehat{\sigma'_0}}{\rhomean}}
& \mbox{where }
\sigma'_0 & = \left. \frac{\partial \sigma(\epsilon,x)}{\partial \epsilon}\right|_{\sigma=\sigma_0}.
\end{align*}


Now consider the case of a shock separating two constant states $q_l,q_r$
in the nonlinear medium, and let $[q]$ denote the jump $q_r-q_l$.  Since it will experience reflections at the 
material interfaces, the shock may conceivably break up into many smaller
discontinuities, as it clearly would in the linear case.  If the shock
is able to persist as a single large discontinuity (due to nonlinear
compression), we might expect that it will also travel at an effective 
velocity; it is natural
to suppose that this velocity will be related to the Rankine-Hugoniot shock
speed in the same way that the effective velocities above are related to
the characteristic speeds.  The R-H jump conditions in a homogeneous medium
give the shock speed
\begin{align} \label{shockspeed}
s = \sqrt{\frac{[\sigma]}{[\epsilon]\rho}}.
\end{align}
Thus it is natural (based on \eqref{ceff}) to consider an "effective shock speed"
\begin{align} \label{effshockspeed}
s_\textup{eff} = \sqrt{\widehat{\left(\frac{[\sigma]}{[\epsilon]}\right)}\frac{1}{\rhomean}},
\end{align}
which we propose as an estimation of the velocity at which the 
largest portion of the jump may travel.  Smaller parts of the shock will 
have spread out due to being reflected more or fewer times.  The fastest-travelling
portion is that which undergoes no reflections.
The strength of this "purely transmitted" 
shock diminishes exponentially in time, so that to a good approximation it travels
at the speed for non-reflected small perturbations in the state ahead of the shock:
$$\hat{c}(\sigma_r) = \left(\int_0^1 \left(\frac{\sigma'_r(x)}{\rho(x)}\right)^{-1/2}dx\right)^{-1}.$$

\subsection{A generalized entropy condition}
A generalized condition for shock stability can be motivated as follows.
The bulk of the shock will advance at speed $s_\textup{eff}$.  If 
$s_\textup{eff}<\hat{c}(\sigma_r)$ then the bulk of the shock will fall behind
the leading (weak) bit.  This will make the main shock weaker, causing it to 
propagate even more slowly, so that more of the strength of the shock will
"escape" ahead of it.  Hence the shock will be converted
into many weak shocks after some time.  

This proposed condition for shock persistence in the periodic medium can be 
viewed as a generalization of the Lax entropy condition for shocks.  Namely, 
characteristics of the corresponding family must impinge on the shock.
In our case, the characteristic speed involved in this condition is the
harmonic average of the sound speed ahead of the shock.
The relevant shock speed is the effective shock speed $s_\textup{eff}$ 
from \eqref{effshockspeed}.

\begin{figure}
\centerline{\includegraphics[width=4in]{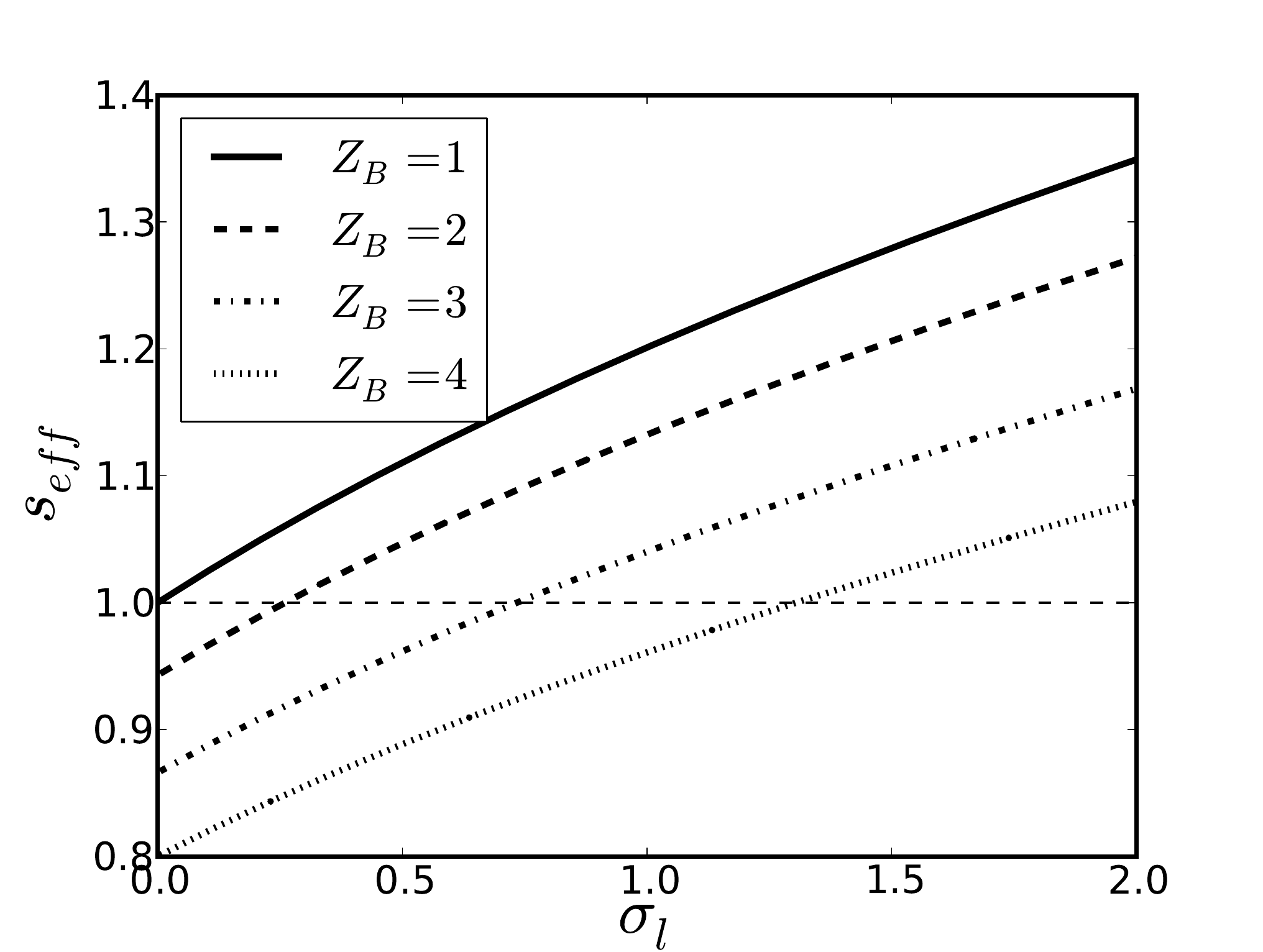}}
\caption{Effective shock speed \eqref{effshockspeed} as a function of stress behind the shock.\label{fig:cshock}}
\end{figure}

The effective shock speed for various values of $Z_B$ (with $\rho_B=K_B=Z_B$)
and a range of stresses 
$\sigma_l$ is plotted in Figure \ref{fig:cshock}.  Here we have taken
$\sigma_r=0$, so that $\hat{c}(\sigma_r)=1$.  Hence we expect shock stability
for a given impedance $Z_B$ and left state $\sigma_l$, if the corresponding 
effective shock speed \eqref{effshockspeed}
is greater than unity.
For the homogeneous medium ($\rho_B=K_B=Z_B=1$),
the effective shock speed is just the ordinary shock speed, and any shock 
is of course predicted to be stable.  However, for heterogeneous media, 
small-amplitude shocks have effective velocity less than one, 
so that effective characteristics do not impinge on the shock from the right.

Based on the foregoing theory, we can define a critical parameter that
we will refer to as the {\em relative effective shock speed}:
\begin{align} \label{eq:seff}
S_\textup{eff} & = \frac{s_\textup{eff}}{\hat{c}(\sigma_r)}.
\end{align}
Notice that $S_\textup{eff}$ depends on the ambient states $\sigma_r,\sigma_l$
as well as the material parameters $\rho(x),\sigma'_r(x)$.  We expect that a
shock will form whenever $S_\textup{eff}$ exceeds unity.

\subsection{Numerical validation}
The true dynamics of a propagating front in the periodic medium
is generally much more complicated than what we have just considered,
since such a front quickly results in many reflected/transmitted shocks and
rarefactions in mutual interaction.  Nevertheless, simulations of
low-contrast media and large-amplitude initial conditions do lead to
persistent large-amplitude shock fronts, whereas simulations of 
higher-contrast media or smaller-amplitude initial conditions such
shocks are not visibly apparent.  For the former case, we have compared directly 
the apparent shock velocity with the predicted value $s_\textup{eff}$.
There is some ambiguity in determining the precise location of the shock,
since the front structure is complex, but in general agreement to within
1-3\% is observed for the range of materials and initial conditions considered
in this work.

\begin{figure}
\centerline{
\includegraphics[width=4in]{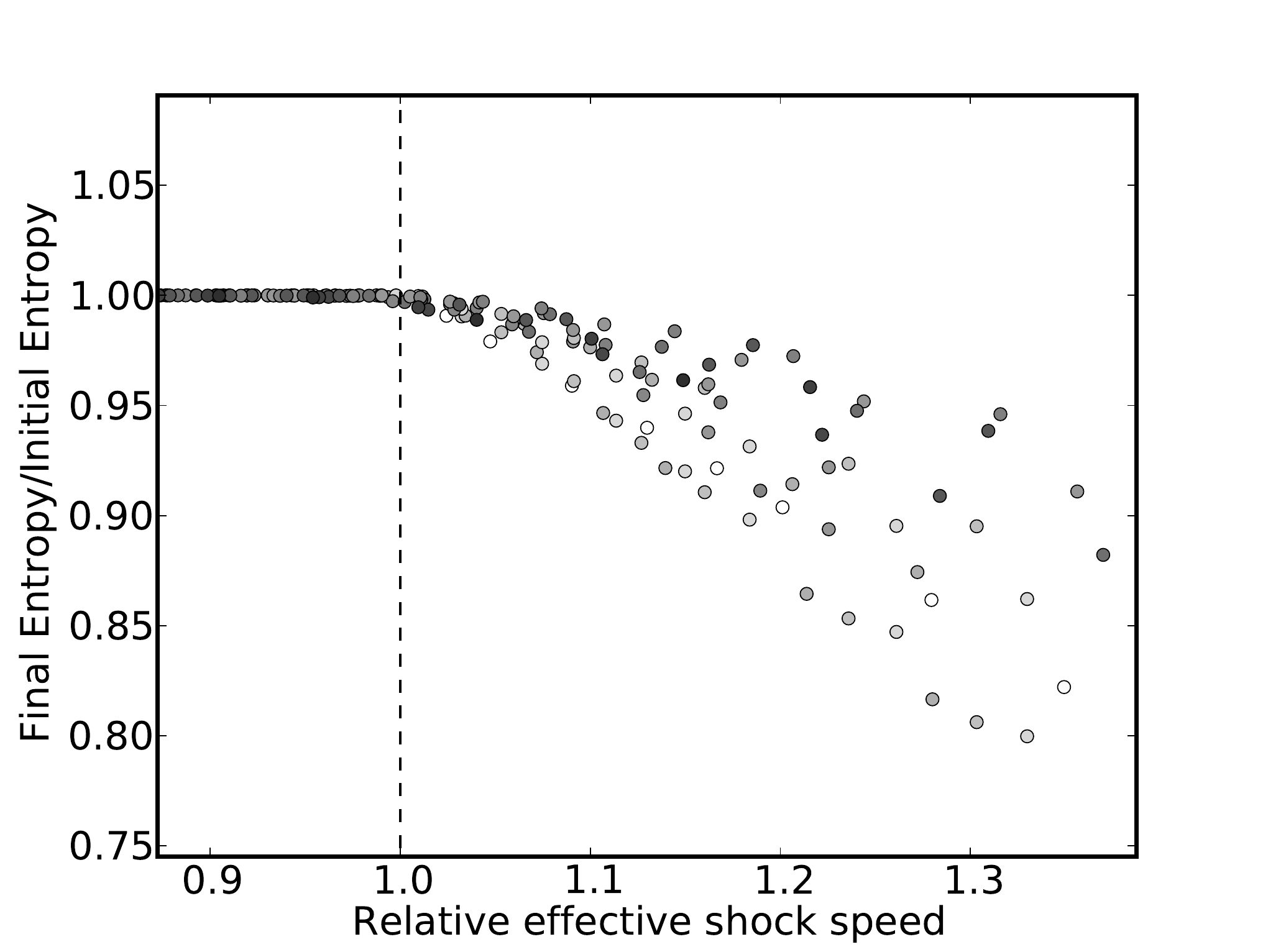}}
\caption{Relative entropy versus relative effective shock speed $S_\textup{eff}$
\eqref{eq:seff} 
for a wide range of materials and initial conditions.  Notice that the 
entropy is near constant for $S_\textup{eff}<1$ (indicating
no shock formation), and decreasing for $S_\textup{eff}>1$ (indicating
shock formation).  The shade of each point indicates the impedance contrast
on a logarithmic scale, with darker points corresponding to higher impedance
contrast.
\label{fig:ent_cshock}}
\end{figure}

We now conduct two tests to quantitatively validate this theory.
In the first, we test whether shocks form from smooth initial conditions.
The initial condition consists of
two uniform initial states separated by a thin smooth transition region.
The solution is evolved to time $t=100$, well beyond
the time for shock formation in a homogeneous medium for all of the initial
states studied.  Then the velocity
is reversed and the simulation continues to $t=200$.  The final entropy 
and initial entropy are compared to determine whether shock formation has
occurred.  In Figure \ref{fig:ent_cshock}, the ratio of final and initial 
entropies is plotted for many tests over the ranges 
$1.5\le\rho_B\le 9,1\le K_B \le 4,0.1\le\sigma_l\le4$, 
and $\sigma_r=0$.  The results are in very good agreement with the theory;
i.e. entropy loss occurs when $S_\textup{eff}>1$
and generally is on the level of numerical error otherwise.  

In the second test, we start with discontinuous initial conditions and
study the entropy evolution in time.  There is some difficulty in choosing
an initial condition.  Using a pure right-going shock in one material quickly 
leads to very strong reflections.  Instead,
we use an initial condition consisting of an "effective shock", where the
left and right states are related by the usual Rankine-Hugoniot conditions
but with the material parameters $\rho,K$ replaced by the effective parameters
$\rhomean,\Kmean$.  Specifically, we take $(\sigma(x,0),u(x,0))=0$ for $x>30$
and $(\sigma(x,0),u(x,0))=(\sigma_l,u_l)$ for $x\le 30$, where
\begin{align*}
u_l & =-\sqrt{\frac{\sigma_l \log(\sigma_l+1)}{\rhomean\Kmean}}.
\end{align*}
This dramatically reduces the severity of initial reflections of the front,
which seems advantageous for our purpose of studying shock propagation.

\begin{figure}
\centerline{
\includegraphics[width=4in]{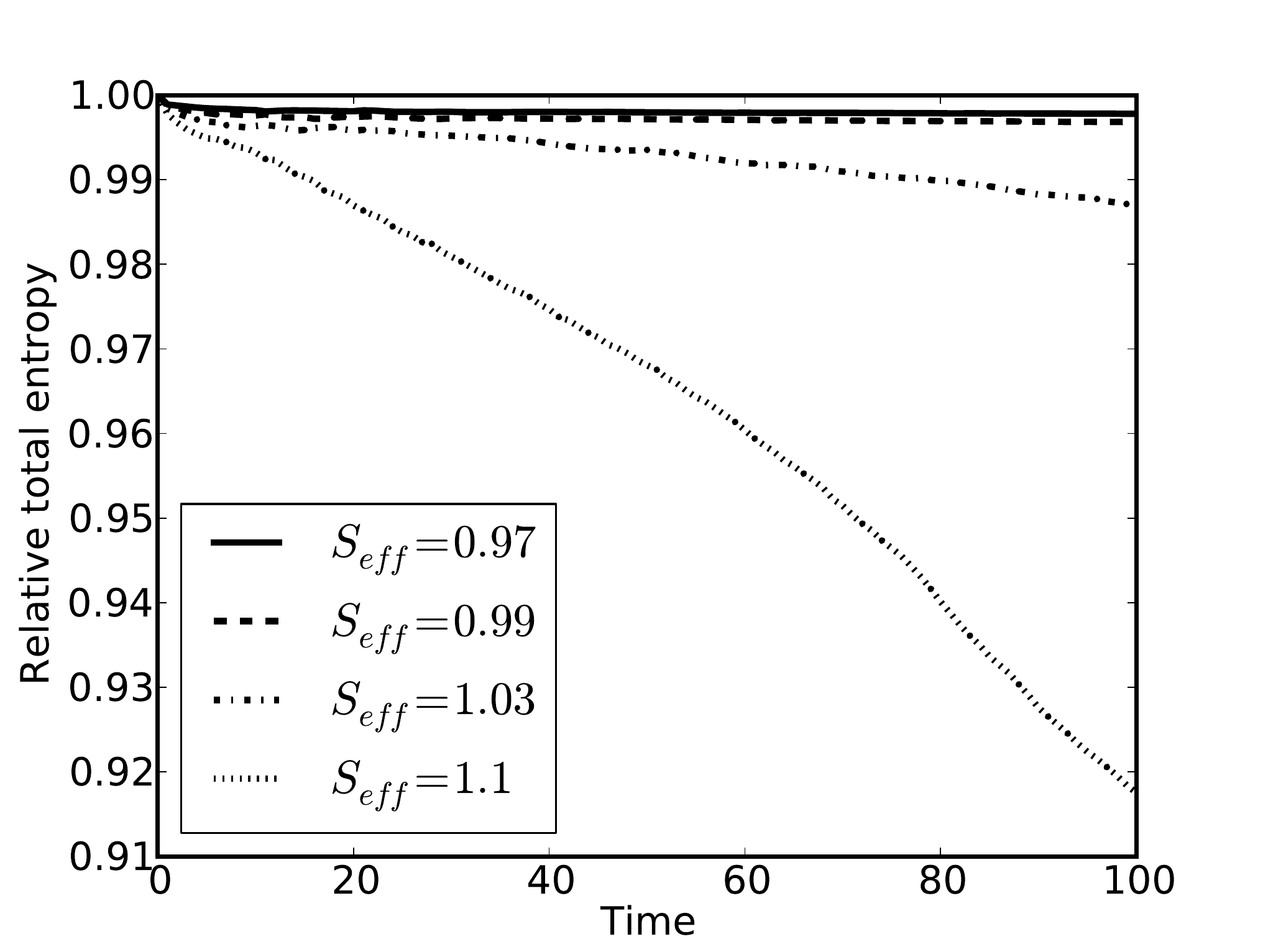}}
\caption{Entropy as a function of time (relative to initial entropy)
for a range of relative effective shock speeds $S_\textup{eff}$ \eqref{eq:seff}.
The medium has $\rho_B=K_B=Z_B=2$.
Notice that the entropy is nearly constant (after a brief initial decrease) for
$S_\textup{eff}<1$, and decreasing for $S_\textup{eff}>1$.\label{fig:ent_cshockic}}
\end{figure}

The total entropy evolution versus time is plotted in Figure \ref{fig:ent_cshockic}
for a range of values of $\sigma_l$ and a medium with $Z_B=\rho_B=K_B=2$.  
The legend indicates the corresponding values
of the relative effective shock speed $S_\textup{eff}$.  In all cases there
is a rapid but small initial entropy loss.  However, for $S_\textup{eff}<1$
the entropy is nearly constant after this initial time.  On the other hand, for
$S_\textup{eff}>1$ the entropy continues to decrease significantly throughout the simulation.

\begin{figure}
\centerline{
\includegraphics[width=4in]{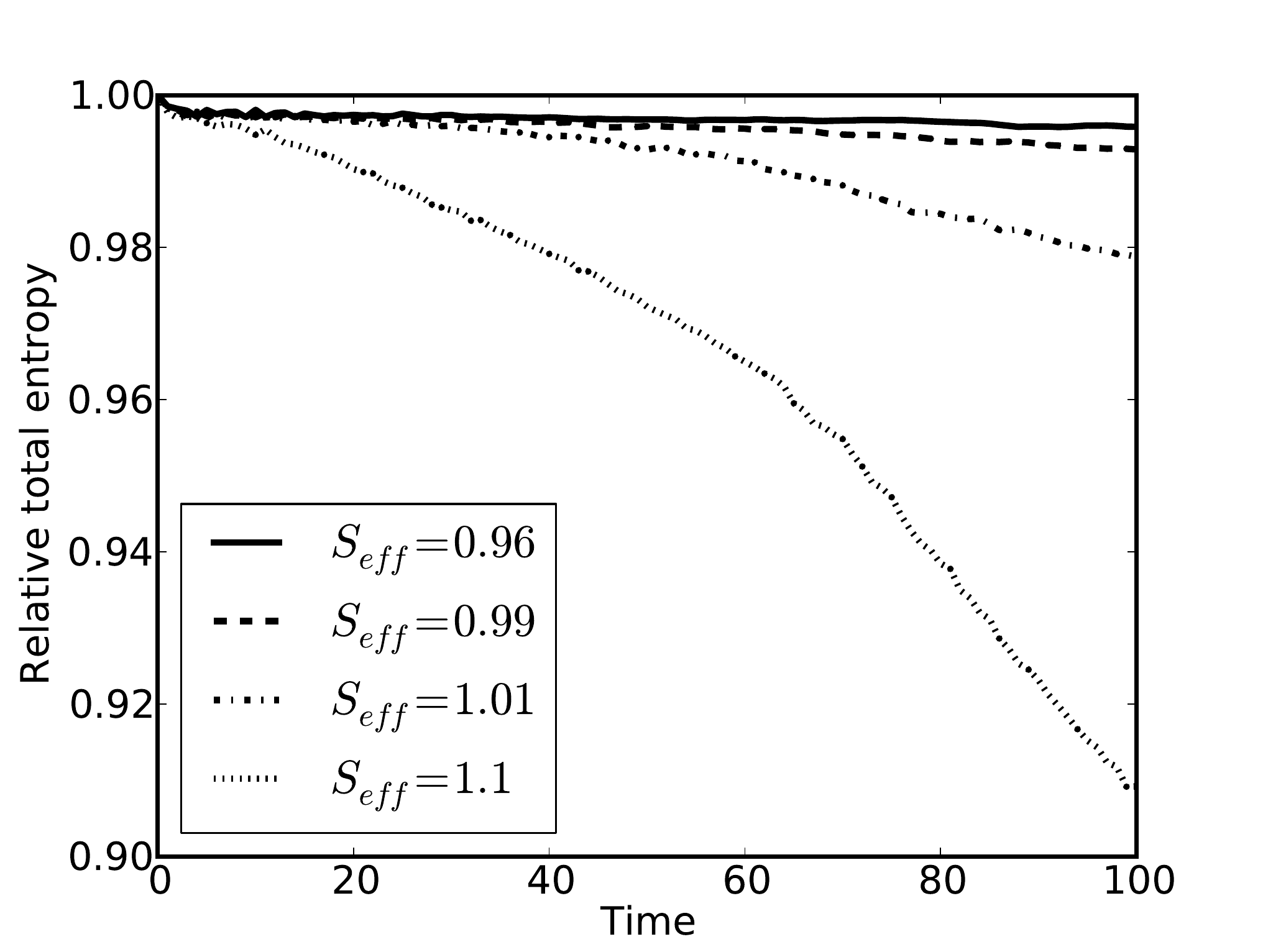}}
\caption{Entropy as a function of time (relative to initial entropy)
for a range of relative effective shock speeds $S_\textup{eff}$ \eqref{eq:seff}.
The medium has $\rho_B=K_B=Z_B=4$.
Notice that the entropy is nearly constant (after a brief initial decrease) for
$S_\textup{eff}<1$, and decreasing for $S_\textup{eff}>1$.\label{fig:ent_cshockic2}}
\end{figure}

Figure \ref{fig:ent_cshockic2} shows similar results for a medium with
$Z_B=\rho_B=K_B=4$.  In this case, we see that for $S_\textup{eff}$ slightly
less than 1, some significant entropy loss continues after the initial
period.  The entropy loss is much more significant when $S_\textup{eff}>1$.

\section{Discussion\label{discussion}}
  The main findings of this work can be summarized as follows:
\begin{enumerate}
  \item Layered periodic materials with varying impedance tend to inhibit 
        shock formation.
  \item Weak shocks in periodic media are unstable, in the sense that they
        do not persist as noticeable discontinuities and do not lead to significant 
        long-term entropy decay.
  \item Initial perturbations with large enough amplitude do lead to shock
        formation and substantial sustained entropy decay.
  \item The formation of shocks can be quantitatively predicted, to a good
        approximation, by a generalized characteristic condition.
\end{enumerate}

In order to detect the formation of shocks in layered periodic media, 
we have studied the time-reversibility and the entropy
evolution of computated solutions.  Based on these experiments and
some intuition, we are led to a simple criterion for shock formation
in terms of one parameter: the relative effective shock speed $S_\textup{eff}$.  
This theory is consistent with a wide range of experiments involving simple 
layered media.  It also seems to be a natural
generalization of the Lax entropy condition, taking into account the 
effective properties of the periodic medium.

This result has potentially important implications both for nonlinear
hyperbolic PDE theory (the fact that shocks appear to be avoided for all
time when $S_\textup{eff}<1$) and for effective medium theory (the fact
that classical shocks may still form in a periodic medium when $S_\textup{eff}>1$).
The result has been formulated in terms of general periodic media and could
be interpreted in a natural way for other heterogeneous (non-periodic)
media.  Whether the same condition for shock formation holds in such 
broader settings is an open question.
Another way in which this theory could be generalized naturally involves
application to more general first-order hyperbolic systems.
Both of these topics are the subject of ongoing research.

A very interesting question touched on in the introduction is that of 
whether it is possible for non-trivial initial conditions to remain smooth 
for all time in the solution of nonlinear hyperbolic PDEs with varying 
coefficients.
Because the approach of the present work has been based on computation and 
observation, it provides tantalizing hints but does not address 
in a strict way the answer to this question.

We finally remark that the interesting behavior of different limiters
with respect to entropy evolution for the waves studied here seems
worthy of further study.



\subsection*{Acknowledgments}
This research was supported in part by NSF grant DMS-0914942 and NIH grant
5R01AR53652-2.

\bibliography{stego}

\end{document}